\documentclass[letterpaper]{aipproc}
\layoutstyle{6x9}
\title[Vertex integrals]{Vertex integrals in heavy-particle
  theories}
\author{Antonio O.\ Bouzas}{
  address={Departamento de F\'{\i}sica Aplicada, Cinvestav-IPN, Apdo.\
           Postal 73 ``Cordemex'', M\'erida 97310, Yucat\'an,
           M\'exico}, 
  email={abouzas@mda.cinvestav.mx},}
\newcommand{\J}[1]{\mathcal{J}_{#1}}
\renewcommand{\H}[1]{\mathcal{H}_{#1}}

\newcommand{\alp}{\alpha_{\scriptscriptstyle +}}
\newcommand{\alm}{\alpha_{\scriptscriptstyle -}}
\newcommand{\scsc}{\scriptscriptstyle}
\newcommand{\loga}[1]{\log\!\left(#1\right)}
\newcommand{\logad}[1]{\log^2\!\left(#1\right)}
\newcommand{\suma}{\sum_{\scsc k,\sigma}}
\newcommand{\zks}{z_{k\sigma}}
\newcommand{\yo}{y_{\scsc 0}}

\newcommand{\Sp}[1]{\mathrm{Li}_{2}\!\left(#1\right)}
\newcommand{\qq}{\mathbf{q}}
\newcommand{\F}[1]{\mathcal{F}_{#1}}
\newcommand{{\zo}}{\ensuremath{z_{\scsc 0}}}
\newcommand{\sgn}{\mathrm{sgn}}
\newcommand{\HH}[1]{\mathsf{H}_{#1}}
\newcommand{\VV}{\mathsf{V}}
\newcommand{\QQ}{\mathsf{Q}}

\begin{document}
\begin{abstract}
  We give the results of complete analytical computations of two- and
  three-point loop integrals ocurring in heavy particle theories, with
  and without velocity change, for arbitrary values of external
  momenta and masses.
\end{abstract}
\maketitle
\section{Introduction}
\label{sec:intro}
In this talk we consider a class of one-loop integrals occurring in
heavy-particle theories \cite{hqet}, with arbitrary real values for
the external masses and residual momenta. We give the results of
complete analytical computations of three-point loop integrals with
and without velocity change, and two-point loop integrals.  The
details of the calculations are given in \cite{bouz}, and in a
forthcoming paper.

\section{Loop integrals}
\label{sec:method}

The loop integrals we consider are of the form, 
\begin{eqnarray}
  \J{2}^{\alpha_1\cdots \alpha_n} & = & \frac{i\mu^{4-d}}{(2\pi)^d}
  \int\! d^d\! \ell \frac{\ell^{\alpha_1}\cdots \ell^{\alpha_n}}{
    \left(2v\!\cdot\! (\ell+k)-\delta M+i\varepsilon\right)
    \left(\ell^2-m^2+i\varepsilon\right)}.  \nonumber\\
  \J{3}^{\alpha_1\cdots \alpha_n} & = & \frac{i\mu^{4-d}}{(2\pi)^d}
  \int\! d^d\! \ell \frac{\ell^{\alpha_1}\cdots \ell^{\alpha_n}}{
    \left(2v_1\!\cdot\! (\ell+k_1)-\delta M_1+i\varepsilon\right)
    \left(2v_2\!\cdot\! (\ell+k_2)-\delta M_2+i\varepsilon\right)
    }  \nonumber\\
  & & \times \frac{1}{\left(\ell^2-m^2+i\varepsilon\right)} \label{eq:ints}\\
  \H{3}^{\alpha_1\cdots \alpha_n} & = & \frac{i\mu^{4-d}}{(2\pi)^d}
  \int\! d^d \ell \frac{\ell^{\alpha_1}\cdots \ell^{\alpha_n}}{
    \left(2v\!\cdot\! \ell-\delta M+i\varepsilon\right)
    \left((\ell-k_1)^2-m^2+i\varepsilon\right)} \nonumber\\
  & & \times \frac{1}{\left((\ell-k_2)^2-m^{\prime
  2}+i\varepsilon\right)} \nonumber
\end{eqnarray}
Here $v_i^\mu$, $i=1,2$, are the velocities of the external heavy
legs, $k_i^\mu$ their residual momenta, and $\delta M_i$ their mass
splittings relative to the common heavy mass of the corresponding
heavy quark symmetry multiplet.  $m$ and $m^\prime$ are the masses of
the light particles within the loops, which in chiral theories
are light pseudoscalar mesons.  These integrals are defined
in $d=4-\epsilon$ dimensions, $\mu$ being the mass scale of
dimensional regularization.  Their degrees of divergence are $n+d-3$
for $\J{2}^{\alpha_1\cdots \alpha_n}$, $n+d-4$ for
$\J{3}^{\alpha_1\cdots \alpha_n}$ and $n+d-5$ for
$\H{3}^{\alpha_1\cdots \alpha_n}$.  The factor of 2 in front of
$v_i^\mu$ corresponds to our normalization of the heavy-particle
propagators.

Our method of calculation \cite{bouz} is to obtain the integrals
(\ref{eq:ints}) as large-mass limits of ordinary loop integrals.  We
closely follow the approach of \cite{thooft} for the computation of
scalar integrals, which are greatly simplified in the large-mass
limit, and the method of \cite{passar} to express tensor integrals in
terms of scalar ones.

\section{Two-point integrals}
\label{sec:2point}

Two-point integrals with one heavy propagator have been given in
\cite{bouz,groz,stew}.  The scalar two-point integral $\J{2}$ is a
function of $m$ and $\Delta=\delta M-2 v\cdot k$.  We write it in
terms of $\xi=\Delta/(2m)$,
\begin{displaymath}
  \label{j2b}
 \J{2}(\Delta,m)  =  \frac{\Delta}{32\pi^2} \left( \frac{2}{\epsilon} +
  \loga{\frac{\overline\mu^2}{m^2}} + 2 \right) + \frac{m}{16\pi^2}
  {\cal F}(\xi) 
\end{displaymath}
with ${\cal F}(x)  =  \sqrt{x^2 - 1 + i\varepsilon} \left[
\loga{x-\sqrt{x^2 - 1 + i\varepsilon}} - \loga{x+\sqrt{x^2 - 1 +
i\varepsilon}}\right]$.  The coefficient of the dimensional
regularization pole vanishes when $\Delta=0$.  This is due to the
fact that the real part of the integrand in $\J{2}$ is parity-odd 
when $\Delta=0.$

The vector two-point integral $\J{2}^{\mu}(v^\alpha,\Delta,m)$ is
given in terms of only one form factor, $ \J{2}^\mu(v^\alpha,\Delta,m)
= F(\Delta,m) v^\mu,$ with $ F(\Delta,m) = v_\mu
\J{2}^\mu(v^\alpha,\Delta,m) = 1/2 A_0(m) + \Delta/2
\J{2}(\Delta,m)~,$ where $A_0$ is the standard one-point scalar
integral (see the appendix of \cite{bouz}).  The second-rank tensor
integral is computed analogously, explicit results being
given in \cite{bouz}.

\section{Three-point integrals with velocity change}
\label{sec:3point}

The scalar three-point integral with velocity change,
$\J{3}=\J{3}(v_1\cdot v_2,\Delta_1,\Delta_2,m)$, where $\Delta_j =
\delta M_j -2 v_j\cdot k_j$, can be expressed in terms of four
dilogarithms \cite{bouz},
\begin{displaymath}
  \J{3}  = \frac{1}{64\pi^2} \frac{1-\Omega^2}{\Omega}
  \left\{\frac{2}{\epsilon} \loga{\alpha} +\logad{\alpha} 
  +\suma (-1)^{k+1} \left[\frac{1}{2} \logad{\frac{-\zks}{\overline\mu}}
      +\Sp{\frac{-\yo}{\zks}}
    \right]\right\}.
\end{displaymath}
The notation is as follows, $\omega = v_1\cdot v_2$, $\Omega =
\sqrt{(\omega-1)/(\omega+1)}$, $\alpha = \omega + \sqrt{\omega^2-1} =
(1+\Omega)/(1-\Omega)$ and $\overline\mu$ is the mass unit in the
$\overline\mathrm{MS}$ scheme.  The sum extends over $k=1,2$ and
$\sigma=\pm$, with
\begin{eqnarray*}
  \yo & = & -\frac{1+\Omega}{2\Omega} \left( \frac{1+\Omega}{2}
    \Delta_1 - \frac{1-\Omega}{2} \Delta_2 \right)\\
  z_{1\pm} & = & \frac{1}{2} \left( \frac{1+\Omega^2}{2\Omega}
    \Delta_1 - \frac{1-\Omega^2}{2\Omega} \Delta_2 \pm
    \sqrt{\Delta_1^2 - 4 m^2 + i \varepsilon} \right) \\  
  z_{2\pm} & = & \frac{\alpha}{2} \left( \frac{1-\Omega^2}{2\Omega}
    \Delta_1 - \frac{1+\Omega^2}{2\Omega} 
    \Delta_2 \pm \sqrt{\Delta_2^2 
    - 4 m^2 + i \varepsilon} \right). 
\end{eqnarray*}
This expression for $\J{3}$ is equivalent to the result given in eq.\ 
(30) of \cite{bouz}, though it has been written in a more compact form
by means of the identity
\begin{displaymath}
  \frac{1}{2} \log^2(z) - \log(z) \log(-z) =- \frac{\pi^2}{2} -
  \frac{1}{2} \log^2(-z), 
\end{displaymath}
valid on the first Riemann sheet of the logarithm, and the identity
(A.2) of \cite{thooft} for the dilogarithm.

In order to compute the vector integral $\J{3}^\mu$ we define two sets
of form factors as $\J{3}^\mu = I_1 v_1^\mu + I_2 v_2^\mu$ and
$F_{1,2} = v_{1,2}\cdot\J{3}$.  These form factors are given by,
\begin{displaymath}
  I_1   =  \frac{1-\Omega^2}{4\Omega^2} \left[-(1-\Omega^2) F_1 +
    (1+\Omega^2) F_2 \right]
    , \hspace{3ex}
  I_2  =  \frac{1-\Omega^2}{4\Omega^2} \left[(1+\Omega^2) F_1 -
  (1-\Omega^2) F_2 \right] ~,
\end{displaymath}
with $F_{1,2} = 1/2 \J{2}(\Delta_{2,1},m) + \Delta_{1,2}/2
\J{3}(v_1\cdot v_2,\Delta_1,\Delta_2,m).$ These equations give an
explicit expression for $\J{3}^\mu$.  For the sake of brevity, we omit
here the results for the tensor integral $\J{3}^{\mu\nu}$, which can
be found in \cite{bouz}.

\section{Three-point integrals with one heavy propagator}
\label{sec:3pointbis}

The scalar three-point integral $\H{3}=\H{3}(v\cdot
q,q^2,\Delta,m,m^\prime)$, with $q=(k_2-k_1)/2$ and $\Delta=\delta M -
v\cdot (k_1+k_2)$, can be expressed in terms of eight dilogarithms as,
\begin{eqnarray*}
  \H{3} & = & \frac{1}{(4\pi)^2}
  \frac{1}{4 |\qq|} \sum_{j=1,2} 
  \left( \F{1} (y_j) + \F{2} (x_j) - \F{3} (z_j)  \right) \\
  \F{1} (x) & = & \Sp{\frac{\zo - 4 |\qq| \alpha}{\zo - 4 |\qq| x}} -
  \Sp{\frac{\zo}{\zo - 4 |\qq| x}} \nonumber \\
  \F{2} (x) & = & - \Sp{\frac{\zo - 4 |\qq| \alpha}{\zo - 4 |\qq| x}}
  - \frac{1}{2} \log^2 \left( \frac{\zo - 4 |\qq| x}{\mu^2}\right)
  \nonumber \\
  \F{3} (x) & \equiv  & \F{1} (x) + \F{2} (x)  = - \Sp{\frac{\zo}{\zo
  - 4 |\qq| x}} - \frac{1}{2} \log^2 \left( \frac{\zo - 4 |\qq|
  x}{\mu^2}\right).
\end{eqnarray*}
The quantities entering these equations are $|\qq|=\sqrt{(v\cdot q)^2
  - q^2}$ ($|\qq|$ is assumed to be real), $\alpha\equiv\alp$ with
$\alpha_\pm=2(v\cdot q\pm|\qq|)$, $\zo = -(m^{\prime 2}-m^2) - \alp
(\Delta-2 |\qq|)$, and,
\begin{eqnarray*}
  \label{eq:roots}
  x_{1,2} & = & v\cdot k_2 + 2 |\qq| -\frac{\delta
    M}{2} \pm \sqrt{\left(v\cdot k_1-\frac{\delta M}{2}\right)^2 - m^2 +
    i \varepsilon} \\
  y_{1,2} & = & \frac{1}{2\alm} \left( 4q^2 + m^{\prime 2} - m^2 \pm
    \sqrt{\left(4 q^2 - (m^\prime + m)^2 \rule{0ex}{2ex}\right)
      \left(4 q^2 - (m^\prime - m)^2 \rule{0ex}{2ex}\right)
      + i \varepsilon\sigma}\right) \\
  z_{1,2} & = & v\cdot k_2 -\frac{\delta M}{2} \pm
    \sqrt{\left(v\cdot k_2 -\frac{\delta M}{2}\right)^2 -
    m^{\prime 2} + i \varepsilon},
\end{eqnarray*}
where in the expression for $y_j$ we denoted $\sigma\equiv \sgn
(q^2)$. $\mu$ is a positive constant with dimension of mass, analogous
to the mass unit in dimensional regularization.  It is not difficult
to show that $\H{3}$ does not depend on $\mu$, it appears there for
purely dimensional reasons.

Tensor three-point integrals $\H{3}^{\alpha_1\cdots\alpha_n} =
\H{3}^{\alpha_1\cdots\alpha_n}(v,k_1,k_2;\delta M,m,m^\prime)$ can be
given in terms of integrals with smaller ranks and fewer points, by
the well-known method of \cite{passar}.  We will consider integrals of
standard form $\HH{3}^{\alpha_1\cdots\alpha_n}(v,q;\Delta,m,m^\prime)
= \H{3}^{\alpha_1\cdots\alpha_n}(v,-q,q;\Delta+v\cdot
(k_1+k_2),m,m^\prime)$
in terms of which we can express $\H{3}$ as,
\begin{displaymath}
  \H{3}^{\alpha_1 \cdots \alpha_n} = \HH{3}^{\alpha_1\cdots \alpha_n}
  (v,q;\Delta,m,m^\prime) + \sum_{j=1}^n r^{\{\alpha_1} \cdots
  r^{\alpha_j} \HH{3}^{\alpha_{j+1}\cdots \alpha_{n}\}}(v,q;\Delta,m,m^\prime),
\end{displaymath}
where $r^\mu\equiv 1/2 (p^\prime+p)^\mu$, and $A^{\{\alpha_1\alpha_2\cdots
  \alpha_s\}} \equiv A^{\alpha_1\alpha_2\cdots\alpha_s} +
  A^{\alpha_2\cdots\alpha_s\alpha_1} + \cdots +
  A^{\alpha_s\alpha_1\cdots\alpha_{s-1}}$.
Clearly, for the scalar integral we have $\H{3}=\HH{3}$.   

For the vector integral we write  
$\HH{3}^\alpha (v,q; \Delta,m_1,m_2) = \VV v^\alpha + \QQ q^\alpha$,
with $\VV=\VV(v\cdot q,q^2,\Delta,m_1,m_2)$ and similarly $\QQ$.  If
$(v\cdot q)^2 -q^2 = |\qq|^2 = 0$, then $q^\alpha\propto v^\alpha$ and
we can set $\QQ=0$.  If $|\qq|^2\neq 0$, 
\begin{displaymath}
  |\qq|^2 \QQ  =  v\!\cdot\! q\, v_\alpha \HH{3}^\alpha - q_\alpha
  \HH{3}^\alpha,
  \quad
  |\qq|^2 \VV  = -q^2 v_\alpha \HH{3}^\alpha - v\!\cdot\! q\, q_\alpha
   \HH{3}^\alpha,
\end{displaymath}
with, 
\begin{eqnarray*}
  v_\alpha \HH{3}^\alpha & = & \frac{1}{2} B_0(4q^2,m_1,m_2) +
  \frac{\Delta}{2} \HH{3}\\
  q_\alpha \HH{3}^\alpha & = & \frac{1}{4} \J{2}(\Delta-2
  v\cdot q, m_2) - \frac{1}{4} \J{2}(\Delta+2 v\cdot q, m_1)
  \nonumber
+\frac{m_1^2-m_2^2}{4} \HH{3}.
\end{eqnarray*}
We have omitted the arguments $(v,q; \Delta,m_1,m_2)$ of
$\HH{3}^\alpha$ on boths sides of these equations for brevity.  $B_0$
is the standard scalar two-point integral, as given in the appendix of
\cite{bouz}.  Higher-rank tensor integrals can be computed analogously.

The results presented in this section were obtained in collaboration
with R.\ Flores Men\-die\-ta.  A detailed derivation will be given
elsewhere.

\section{Acknowledgements}

This work has been partially supported by Conacyt and SNI.

\end{document}